\documentclass{article}

\usepackage{arxiv}

\usepackage[utf8]{inputenc} 
\usepackage[T1]{fontenc}    
\usepackage{hyperref}       
\usepackage{url}            
\usepackage{booktabs}       
\usepackage{amsfonts}       
\usepackage{nicefrac}       
\usepackage{microtype}      
\usepackage{lipsum}
\usepackage{graphicx}

\title{Gaze Patterns Predict Preference and Confidence in Pairwise AI Image Evaluation\thanks{© 2026 Copyright held by the owner/author(s). This is the author's version of the work. It is posted here for your personal use. Not for redistribution. The definitive Version of Record was published in \textit{2026 Symposium on Eye Tracking Research and Applications (ETRA '26)}, \url{https://doi.org/10.1145/3797246.3803031}}}

\author{
 Nikolas Papadopoulos \\
  Columbia University\\
  New York, NY\\
  \texttt{np2832@columbia.edu} \\
   \And
 Shreenithi Navaneethan \\
  Columbia University\\
  New York, NY\\
  \texttt{sn3144@columbia.edu} \\
  \And
 Sheng Bai \\
  Columbia University\\
  New York, NY\\
  \texttt{sb5019@columbia.edu} \\
    \AND
 Ankur Samanta \\
  Columbia University\\
  New York, NY\\
  \texttt{as7416@columbia.edu} \\
    \And
 Paul Sajda \\
  Columbia University\\
  New York, NY\\
  \texttt{psajda@columbia.edu} \\
}

\begin{document}
\maketitle
\begin{abstract}
Preference learning methods, such as Reinforcement Learning from Human Feedback (RLHF) and Direct Preference Optimization (DPO), rely on pairwise human judgments, yet little is known about the cognitive processes underlying these judgments. We investigate whether eye-tracking can reveal preference formation during pairwise AI-generated image evaluation. Thirty participants completed 1,800 trials while their gaze was recorded. We replicated the gaze cascade effect, with gaze shifting toward chosen images approximately one second before the decision. Cascade dynamics were consistent across confidence levels. Gaze features predicted binary choice (68\% accuracy), with chosen images receiving more dwell time, fixations, and revisits. Gaze transitions distinguished high-confidence from uncertain decisions (66\% accuracy), with low-confidence trials showing more image switches per second. These results show that gaze patterns predict both choice and confidence in pairwise image evaluations, suggesting that eye-tracking provides implicit signals relevant to the quality of preference annotations.
\end{abstract}


\section{Introduction}
The alignment of artificial intelligence (AI) systems with human values increasingly depends on learning from human preferences. Reinforcement Learning from Human Feedback (RLHF) has become foundational for aligning language models, with approaches spanning scalar reward signal methods \cite{rlhf-ouyang-2022} and direct preference optimization (DPO) \cite{dpo-rafailov-2023}. These methods have recently expanded to text-to-image generation \cite{ddpo_black_2023, dpok-fan-NEURIPS2023_fc65fab8, prabhudesai2023aligning, draft-clark-2024-iclr, diffusionDPO-Wallace_2024_CVPR, DSPO-ICLR2025_7f70331d, lee2023aligning}, supported by large-scale preference datasets \cite{pickapic-kirstain-2023, hpsv2-wu-2023, xu2023imagereward, richhf2024, MPS-zhang-2024-cvpr}. At the core of these methods lies a fundamental primitive: presenting humans with two options and collecting binary preference judgments. While these pairwise comparisons directly shape model behavior through reward modeling or policy optimization, little is known about the cognitive processes by which humans form such preferences during evaluation.

Preference learning pipelines observe only the final choice, not the decision process that produced it. The same label can arise from confident deliberation or uncertain guesswork. Although some datasets collect self-reported confidence, such annotations are often discarded during training \cite{Bai2022ConstitutionalAH} and may not fully capture the underlying cognitive dynamics \cite{NisbettWilson1977}. Inter-annotator agreement measures consistency across annotators but does not reveal reliability within individual judgments \cite{rlhf-ouyang-2022, casper2023open}. Recent work has begun incorporating confidence signals, whether estimated from model outputs \cite{huang-etal-2024-enhancing-language, gdpo-2024, wang2024secretsrlhflargelanguage} or inferred from response time \cite{sawarni-2025-response-time}, but model-based approaches do not observe human behavior, and response time provides only a single aggregate measure per trial. Eye-tracking offers richer information: gaze patterns can reveal how humans allocate attention, compare options, and form preferences in real time.

Eye-tracking also provides a window into preference formation. A robust finding in decision-making is the gaze cascade effect: during pairwise comparison, visual attention progressively shifts toward the option that is ultimately chosen in the final second before the response \cite{Shimojo2003-gaze-cascade, Simion-shimojo-2006, Simion-shimojo-2007InterruptingCascade}. Gaze not only tracks emerging preferences but can also influence their formation \cite{Armel_Beaumel_Rangel_2008}, a pattern replicated across diverse tasks \cite{Glaholt01112009, Schotter01092010, Pärnamets-2015-gazecascade-moral}. However, gaze dynamics have not been examined in the context of AI-generated content evaluation, where preference data for alignment is collected. Understanding how preferences form in this setting could shed light on the reliability of preference annotations and reveal implicit signals of decision confidence.

Recent work has begun exploring eye-tracking as a signal for AI alignment, with a focus on text-based evaluation tasks. \cite{Lopez-Cardona-2025-Seeing-eye, galliamov-etal-2025-enhancing-rlhf-humna-gaze} incorporated synthetic gaze features into reward models for RLHF-style pipelines, while \cite{Lopez-Cardona2025-oasst-etc} introduced OASST-ETC, an eye-tracking corpus capturing reading behavior during sequential evaluation of large language model responses. \cite{kiegeland2024_pupil_becomes_master} showed that eye-tracking can be used to construct preference datasets for DPO in sentiment generation, and \cite{bondar-etal-2025-aleyegnment} augmented DPO with gaze-based loss terms for linguistic acceptability classification. Across these studies, gaze is either synthetically generated or recorded during text-based reading tasks, where participants evaluate individual responses in isolation or view alternatives sequentially, with preference labels constructed offline or obtained after serial presentation.

In this study, we extend eye-tracking-based alignment research to text-to-image preference evaluation. Unlike prior work that records gaze during sequential or single-response evaluation, we capture gaze during simultaneous pairwise comparison: participants view a text prompt alongside two AI-generated images simultaneously, indicate their preference, and report decision confidence. This design enables observation of the real-time dynamics of preference formation as decisions unfold. This work also extends classic gaze-cascade paradigms to text-to-image evaluation, a task that requires evaluators to integrate multiple criteria: prompt alignment, aesthetic quality, visual coherence, and the presence of artifacts.

\begin{figure}
    \centering
    \includegraphics[width=0.7\linewidth]{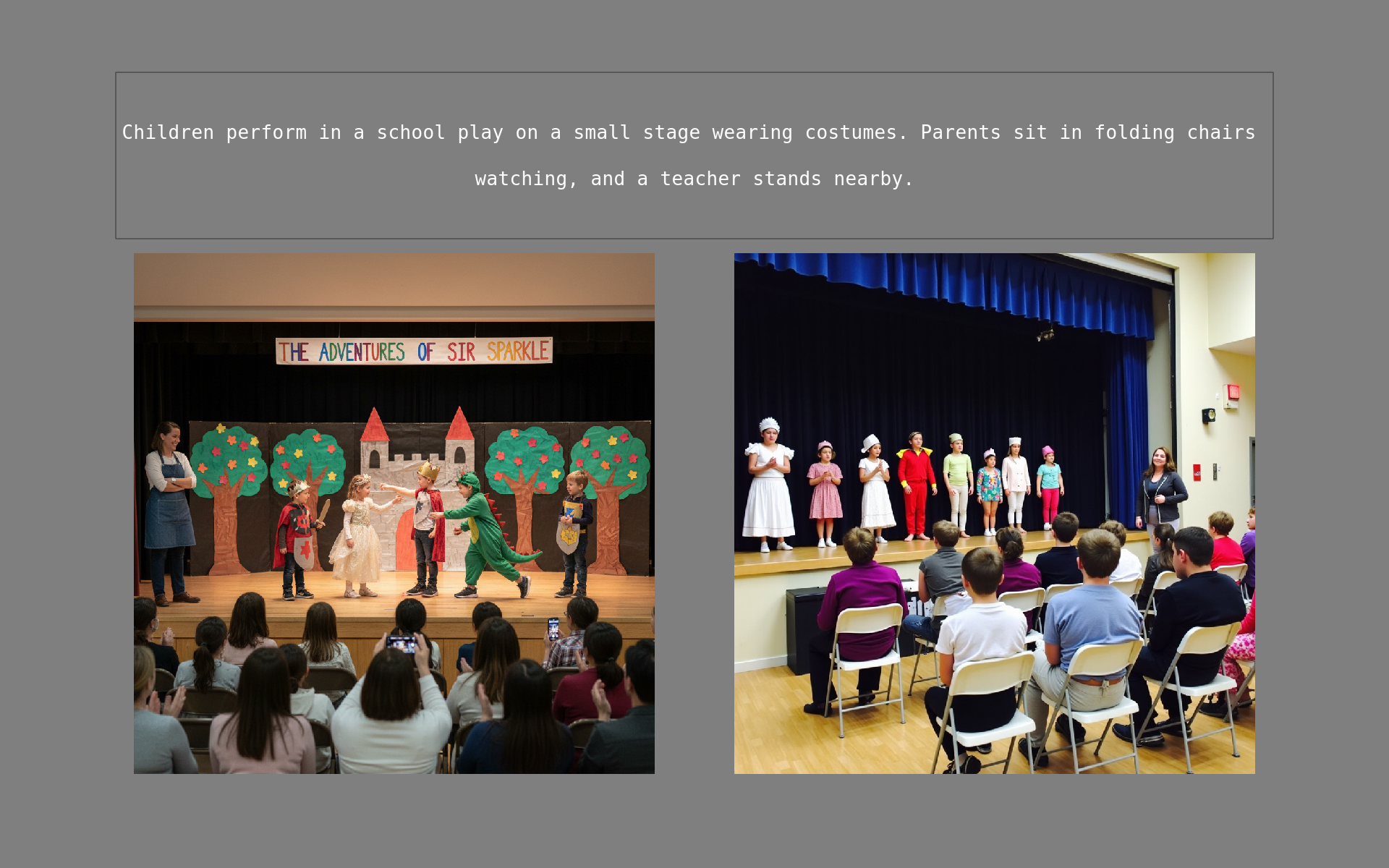}
    \caption{Example trial stimulus.}
    \label{fig:example_stimuli}
\end{figure}

Our key contributions include:

\begin{itemize}
    \item \textbf{Cognitive validation}: We show that the gaze cascade effect extends to multi-attribute evaluation of AI-generated images, with gaze shifting toward chosen images $\sim$1 second before decision.
    \item \textbf{Preference prediction}: Gaze features predict binary preference choice above chance (68\% accuracy), with chosen images receiving greater dwell time, fixations, and revisits, indicating that preference-related information is present throughout the decision process
    \item \textbf{Confidence prediction}: Gaze transition rate predicts decision confidence (66\% accuracy), with low-confidence decisions involving more frequent switching between options, while cascade dynamics remain unchanged across confidence levels. This suggests that gaze transitions capture decision uncertainty and may serve as a signal for identifying less reliable annotations.
\end{itemize}

\section{Methods}
\subsection{Experimental Setup}

\paragraph{\textbf{Task Overview.}} We designed a pairwise image evaluation task that mirrors the RLHF/DPO preference annotation structure. Participants viewed a text prompt alongside two AI-generated images and indicated their preference using arrow keys while their eye movements were tracked (Fig. ~\ref{fig:example_stimuli}). Participants were not given explicit evaluation criteria, reflecting real-world annotation scenarios where evaluators apply their own judgment. Each trial was self-paced with no time constraints. Following each choice, participants rated their confidence (low, medium, or high).

\paragraph{\textbf{Procedure.}} Each participant completed 60 trials divided into two parts of 30, with an optional break between parts. Three practice trials preceded the main experiment. Each trial began with a central fixation cross (2 s), followed by simultaneous presentation of the prompt and both images. Participants pressed the left or right arrow key to indicate their preference, then rated their confidence using the arrow keys: Left (low), Down (medium), or Right (high). Participants were encouraged to maintain attention and evaluate each pair carefully.

\paragraph{\textbf{Stimulus Assignment.}} We compiled a pool of 150 prompt-image pairs from our dataset sources (see Section~\ref{sec:section-image-stimuli}). Each participant viewed a randomly selected subset of 60 pairs, ensuring that each pair was annotated by at least 10 different participants. For each trial, the left-right positioning of images was randomly determined to eliminate spatial response biases.

\paragraph{\textbf{Participants.}} 30 individuals participated in our study (17 female, 13 male, mean age = 24.7 $\pm$ 2.8). All participants provided written consent in a manner approved by the Columbia University Institutional Review Board (IRB). Participants received 20\$ for their participation. Participants had normal or corrected-to-normal vision, and none reported neurological or psychiatric histories or medications.

\paragraph{\textbf{Apparatus.}} The stimuli were displayed on a 24.1-inch monitor (1920 x 1200 resolution) at a viewing distance of approximately 60cm (without chinrest). The experiment was implemented using PsychoPy \cite{Peirce2019PsychoPy2}, with eye-tracking data synchronized via Lab Streaming Layer (LSL) \cite{Kothe2025LSL}. Eye movements were recorded using Tobii Pro Fusion at a sampling rate of 250Hz. A 9-point calibration procedure was performed at the beginning of each session using Tobii Pro Eye Tracker Manager software. Calibration accuracy was verified using a custom 7-point validation covering the experimental display regions (prompt box, left image, right image). Validation was conducted at the beginning of each part. Mean validation accuracy was 0.76° (Part A: 0.75° $\pm$ 0.28°; Part B: 0.77° $\pm$ 0.21°), with 83\% of validations $\le$ 1.0

\paragraph{\textbf{Eye-Tracking Data Processing.}} Binocular gaze position was recorded, with gaze coordinates computed as the average of both eyes. Fixations were extracted using the Dispersion-Threshold Identification (I-DT) algorithm \cite{i-dt-algorithm-2000} with a maximum dispersion of 1.0° and minimum duration of 100 ms. Fixations were mapped to three Areas of Interest (AOIs): left image, right image, and prompt region. For each trial, we extracted 23 gaze features across five categories (Table~\ref{tab:gaze_features}).

\begin{table}[h]
\centering
\caption{Gaze feature categories.}
\label{tab:gaze_features}
\begin{tabular}{ll}
\hline
\textbf{Category} & \textbf{Features} \\
\hline
Per-AOI metrics      & Dwell time, fixation count, mean fixation \\
                     & duration, revisit count \\
Image comparisons    & Dwell time ratio, fixation ratio, absolute \\
                     & differences between images \\
Temporal markers     & First/last fixation location, time to \\
                     & first image fixation \\
Gaze transitions     & Shifts between images, between each \\
                     & image and prompt, total transitions \\
Fixation variability & SD of fixation durations \\
\hline
\end{tabular}
\end{table}

\subsection{Image Stimuli}
\label{sec:section-image-stimuli}
We curated 150 prompt-image pairs from three sources to represent the diversity of modern AI-generated image evaluation. Each trial consisted of one text prompt and two AI-generated candidate images depicting photorealistic real-world scenarios, including everyday objects, people, environments, and activities.

\paragraph{\textbf{Open-Source Datasets}} We included 60 pairs from \textit{Open-image-preferences} \cite{berenstein2024-openpreferencesdataset} (images generated using FLUX.1 Dev and Stable Diffusion 3.5 Large) and 40 pairs from \textit{Rapidata} \cite{rapidata-christodoulou2024findingsubjectivetruthcollecting} (images generated using DALL-E 3, Flux.1 Pro, Midjourney 5.2, and Stable Diffusion 3). Both datasets contained pre-existing human preference annotations. We applied quality filtering to exclude trials with severe artifacts, ensuring that evaluation focused on subtle quality differences rather than obvious defects.

\paragraph{\textbf{Custom-Generated Pairs}} We created 50 additional pairs using complex, detailed prompts depicting daily life scenes with rich descriptive detail and contextual elements (Fig. ~\ref{fig:example_stimuli}). For each prompt, we generated images using Google Gemini, DALL-E 3, and Flux.1 Dev, then manually selected the two highest-quality outputs to form each trial. This ensured that custom trials posed challenging evaluations, with subtle differences in interpretation, style, or execution across high-quality generations.

\section{Results}

\subsection{Gaze Cascade Analysis}
\label{sec:gaze_cascade_analysis}
We analyzed the likelihood of fixating on the eventually chosen image as a function of time before the decision (i.e., the key press response) to test whether the gaze cascade effect \cite{Shimojo2003-gaze-cascade} replicates in AI image evaluation. Trials were aligned at the moment of response and analyzed backward in time. The analysis window spanned the final 4~s before decision, determined by subtracting one standard deviation from the mean response time ($M = 11.0$~s, $SD = 6.6$~s, pooled across trials) to ensure adequate data coverage across trials. This window was divided into 120 bins (33.3~ms per bin). For each bin, we computed the proportion of fixation time directed toward the chosen image relative to total image-viewing time. Fixations on other regions were excluded, and bins with no valid image fixations were treated as missing. For each participant, bin-wise likelihoods were averaged across trials to obtain a single likelihood curve. The grand average and 95\% confidence intervals were then computed across participants. A sigmoid function (with four parameters: starting level, elevation, inflection point, and slope) was fit to the grand average. Parameters were estimated using nonlinear least-squares optimization, with model fit assessed using $R^2$. To examine whether cascade dynamics differed by decision confidence, we separately analyzed low-, medium-, and high-confidence trials. We fit both sigmoid and piecewise linear models to each confidence level. The piecewise linear model captures two interpretable parameters: the breakpoint (when gaze begins shifting from chance toward the chosen image) and the slope (the rate of preference accumulation).

\begin{figure}[h!]
    \centering
    \includegraphics[width=0.8\linewidth]{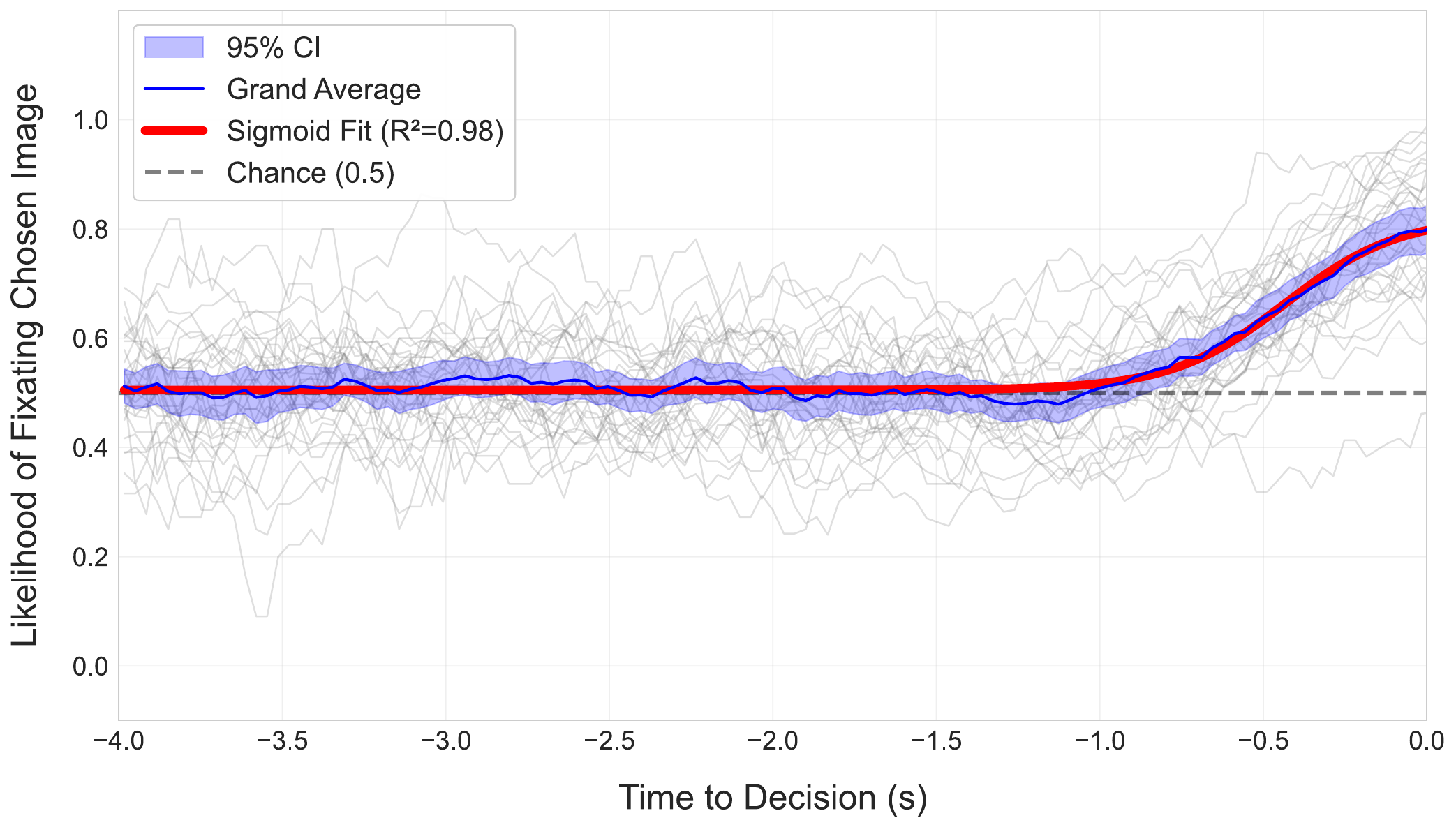}
    \caption{Gaze cascade effect in AI image evaluation. Likelihood of fixating on the eventually chosen image as a function of time before decision (key press). The blue line shows grand average across participants (N=30), with individual participants shown in gray lines.}
    \label{fig:cascade_effect}
\end{figure}

Figure~\ref{fig:cascade_effect} illustrates the gaze cascade effect across all trials. Participants’ gaze was initially distributed evenly between the two images (near 50\% chance level) but progressively shifted toward the eventually chosen image in the final second before the decision. The sigmoid model provided an excellent fit to the data ($R^2 = 0.98$). By the moment of decision, participants fixated on the chosen image approximately 80\% of the time. Although individual participant curves exhibited substantial variability in magnitude, the overall pattern was highly consistent: nearly all participants showed the characteristic transition from balanced attention to preferential gaze toward the selected image during the pre-decision period.

\begin{figure}[h!]
    \centering
    \includegraphics[width=0.9\linewidth]{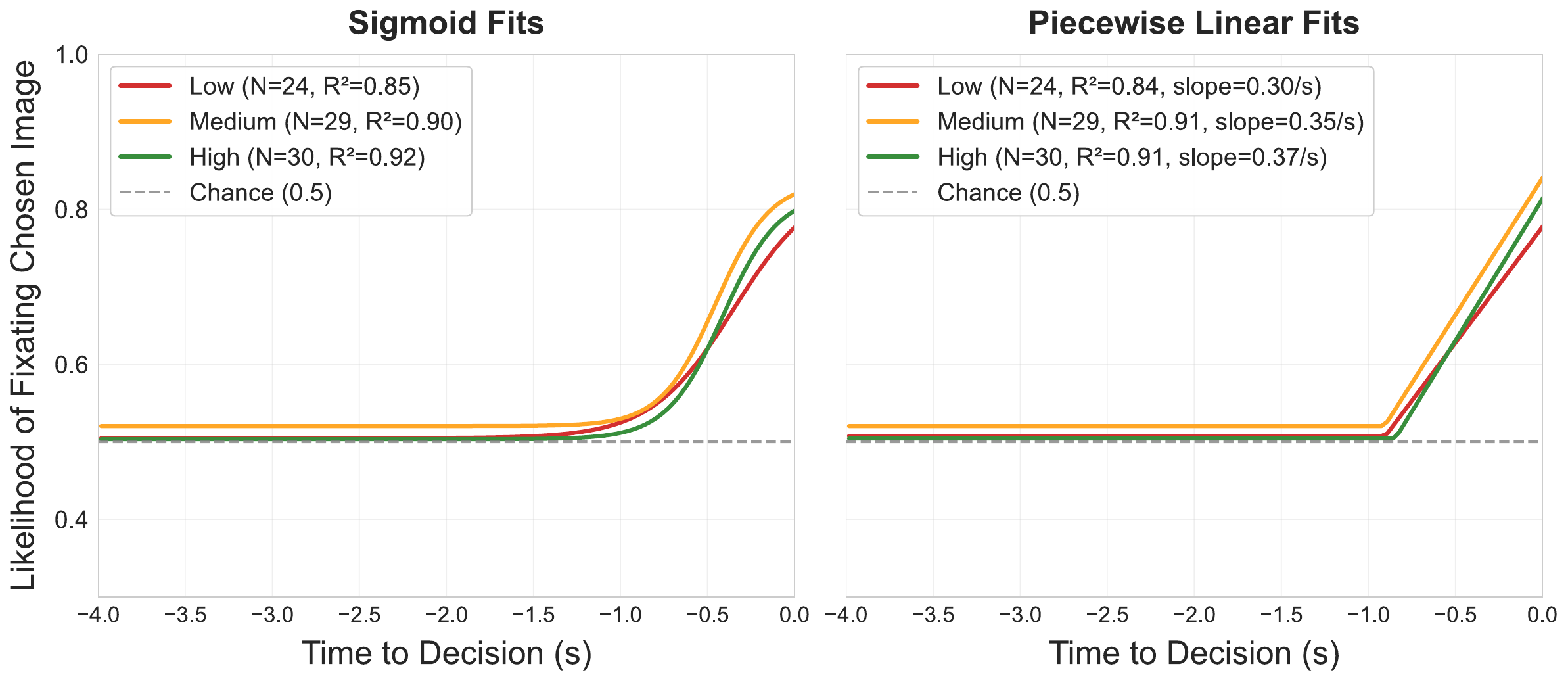}
    \caption{Gaze cascade dynamics by confidence level.}
    \label{fig:cascade_effect_by_confidence}
\end{figure}

To examine whether cascade dynamics vary with decision confidence, we stratified trials by self-reported confidence and fit sigmoid and piecewise linear models to each condition separately (Fig. ~\ref{fig:cascade_effect_by_confidence}). Participants with fewer than 5 trials in a given confidence level were excluded from that condition's analysis to ensure reliable curve estimation, resulting in sample sizes of N=24 (low), N=29 (medium), and N=30 (high). The gaze cascade effect was present across all confidence levels, with sigmoid fits achieving $R^2$ values of 0.85 (low), 0.90 (medium), and 0.92 (high). Low-confidence trials showed numerically shallower piecewise slopes (0.30/s vs 0.35–0.37/s) and reached lower peak likelihood at decision time (78\% vs 82–84\%), but bootstrap analysis (10,000 iterations) revealed no statistically significant differences in any model parameters (slope, inflection point, breakpoint, or asymptote) between confidence levels (all 95\% CIs included zero). This suggests that the progressive allocation of gaze toward the chosen image reflects a stable mechanism of preference formation regardless of decision certainty.

\subsection{Predicting Preference from Gaze Features}
\label{sec:preference_prediction}

\begin{table}[h]
\centering
\caption{Gaze metrics between chosen and unchosen images. Paired t-tests on participant-level means (N=30)}
\label{tab:gaze_metrics}
\begin{tabular}{lcccccc}
\hline
\textbf{Metric} & \textbf{Chosen} & \textbf{Unchosen} & \textbf{$p$} & \textbf{Cohen $d$} \\
\hline
Dwell time (s)      & 2.30 & 2.13 &  $<.001$ & 0.99 \\
Fixations           & 10.7 & 9.7  &  $<.001$ & 1.16 \\
Revisits            & 3.3  & 2.9  &  $<.001$ & 1.77 \\
\hline
\end{tabular}
\end{table}

To assess whether gaze features could predict choice (left vs.\ right image selected), we trained classifiers on the full dataset of 1,800 trials from 30 participants. The dataset was balanced by design (left: 50.1\%, right: 49.9\%), eliminating position bias as a confound. Features were computed relative to spatial position (left/right) rather than choice outcome, as position-based features are available in real-time prediction scenarios where choice is unknown. We evaluated logistic regression, Random Forest, and XGBoost using 5-fold group cross-validation with participants as the grouping variable. Logistic regression performed best and is reported. We first tested all gaze features, including last fixation location, then excluded last fixation to assess whether trial-level gaze features (dwell time, fixation counts, etc.) provide predictive information beyond final gaze position.

\begin{figure}[h!]
    \centering
    \includegraphics[width=0.7\linewidth]{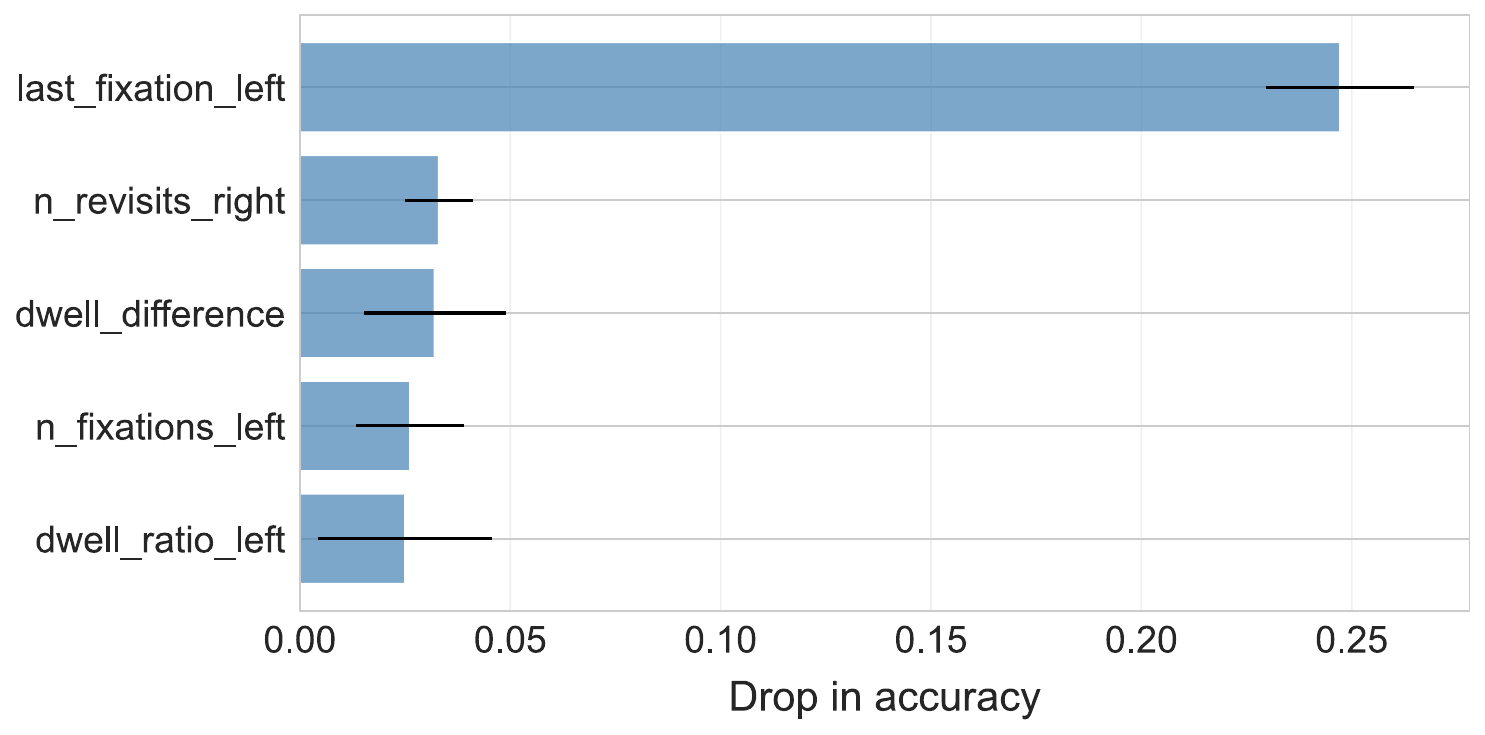}
    \caption{Permutation importance for preference prediction (top 5 features). Error bars show SD across 5 folds.}
    \label{fig:permutation_importance}
\end{figure}

With all gaze features, logistic regression achieved 75.1\% accuracy (ROC-AUC = 0.79). Permutation importance analysis revealed that the last fixation location dominated prediction, approximately 7x greater than any other feature (Fig.~\ref{fig:permutation_importance}). A model using last fixation alone achieved 76\% accuracy, indicating that other features provided minimal additional signal. The first fixation location showed no predictive value (accuracy at chance level), indicating that preference-related gaze patterns emerge during the decision process rather than reflecting initial orienting biases.

Excluding last fixation, accuracy dropped to 67.6\% (ROC-AUC = 0.71), still well above the 50\% baseline (95\% CI [61.2\%, 74.0\%]). This suggests that gaze patterns favoring the chosen image develop throughout the trial, not only at the final moment before response. Paired $t$-tests on participant-level means (Table~\ref{tab:gaze_metrics}) confirmed this pattern: participants dwelled longer on the chosen image ($M = 2.30$s vs.\ $2.13$s, $t(29) = 5.43$, $p < .001$, $d = 0.99$), made more fixations ($M = 10.7$ vs.\ $9.7$, $t(29) = 6.36$, $p < .001$, $d = 1.16$), revisited it more frequently ($M = 3.3$ vs.\ $2.9$, $t(29) = 9.67$, $p < .001$, $d = 1.77$). 

\subsection{Predicting Confidence from Gaze Features}

We first examined whether participants systematically differed in reporting confidence levels using Wilcoxon signed-rank tests with Bonferroni correction ($\alpha_{crit} = .017$). Participants exhibited a bias toward reporting high confidence. High-confidence responses ($M = 28.0$ per participant) were significantly more frequent than low-confidence responses ($M = 9.9$; Wilcoxon $Z = 4.69$, $p < .001$), as were medium-confidence responses ($M = 22.1$) compared to low ($Z = 5.83$, $p < .001$). The difference between high and medium confidence was not significant ($Z = 1.64$, $p = .050$).

\begin{table}[h]
\centering
\caption{Confidence prediction performance. Image transitions refer to gaze shifts between the two images. All models use logistic regression with group 5-fold cross-validation ($N = 1{,}740$ trials, 29 participants).}
\label{tab:confidence_prediction}
\begin{tabular}{lcc}
\hline
\textbf{Features} & \textbf{Accuracy} & \textbf{ROC-AUC} \\
\hline
Image transitions      & $66.1 \pm 5.1$\% & 0.70 \\
All gaze features      & $62.4 \pm 4.3$\% & 0.67 \\
Response time          & $58.1 \pm 1.2$\% & 0.66 \\
Baseline (majority)    & 55.0\%           & ---  \\
\hline
\end{tabular}
\end{table}

To assess whether gaze features could predict decision confidence, we trained classifiers to distinguish high-confidence trials from uncertain trials (low and medium confidence combined). This binary framing reflects the practical goal of identifying high-quality preference data for model training. One participant was excluded due to insufficient variance in confidence ratings (59/60 trials rated high confidence), yielding 1,740 trials from 29 participants. The resulting dataset was approximately balanced (high: 45\%, uncertain: 55\%). We used the same classification setup as preference prediction (Section~\ref{sec:preference_prediction}). Logistic regression performed best and is reported.

Single-feature analysis revealed that the number of transitions between the two images was the strongest predictor of confidence (accuracy $= 66.1\%$, ROC-AUC $= 0.70$), outperforming response time alone ($58.1\%$, ROC-AUC $= 0.66$) and even the full feature set ($62.4\%$, ROC-AUC $= 0.67$) (Table~\ref{tab:confidence_prediction}). The finding that additional features degraded rather than improved performance suggests most gaze features provide redundant information, primarily reflecting trial duration rather than distinct aspects of decision processing.

To determine whether this effect reflected genuine differences in comparison behavior or merely trial duration, we analyzed transition rate (transitions between images per second) across confidence levels using a linear mixed-effects model with confidence level as a fixed effect and participant as a random intercept. Low-confidence trials showed significantly higher transition rates ($M = 0.38$/s) than medium ($M = 0.34$/s) and high-confidence ($M = 0.31$/s) trials. The model confirmed significant differences for both high vs.\ low ($\beta = -0.066$, 95\% CI $[-0.089, -0.042]$, $z = -5.50$, $p < .001$) and medium vs.\ low ($\beta = -0.040$, 95\% CI $[-0.063, -0.017]$, $z = -3.40$, $p < .001$).

Results suggest that low confidence reflects more frequent switching between options rather than simply longer deliberation. This contrasts with the cascade analysis, which revealed no significant differences across confidence levels (see Section~\ref{sec:gaze_cascade_analysis}). The likelihood curve measures the proportion of gaze on the chosen image at each moment, but cannot capture switching frequency. Confidence affects this exploration behavior, i.e., how often participants alternate between options to compare them.

\section{Conclusion}
This study examined whether eye-tracking reveals preference formation during pairwise evaluation of AI-generated images, a task equivalent to RLHF/DPO annotation workflows. Three main findings emerged. First, the gaze cascade effect, well-established in decision-making research, extends to multi-attribute evaluation of AI-generated images, with gaze shifting toward the chosen image approximately one second before the behavioral response. Second, gaze features predicted binary choice with 68\% accuracy, driven by cumulative attentional patterns: chosen images received more dwell time, fixations, and revisits throughout the decision process. Third, gaze transition rate, but not cascade dynamics, predicted decision confidence (66\% accuracy), with low-confidence trials showing more frequent switching between images.

Several limitations should be noted. For confidence prediction, we combined low and medium confidence into a single "uncertain" category to create a balanced classification task; this simplification discards potentially useful granularity in the confidence signal. Additionally, our analysis focused on aggregate gaze patterns at the image level and did not examine fixations on specific regions or semantic content, for instance, whether participants fixate on artifacts or on semantically important regions that match the prompt.

Future work should address these limitations and extend toward practical integration with preference learning pipelines. We consider several pathways for incorporating gaze into reward model training. The first is reward model augmentation, where gaze features serve as auxiliary inputs alongside binary preference labels. Existing work has explored this using synthetic gaze \cite{galliamov-etal-2025-enhancing-rlhf-humna-gaze, Lopez-Cardona-2025-Seeing-eye}, though such signals do not capture actual human attention. A second approach is implicit preference learning, training reward models purely from gaze without explicit labels. However, our findings suggest this may be premature: preference prediction accuracy is modest (68\%), and the cascade effect emerges only one second before response, offering limited advantage over simply collecting explicit feedback.

The most actionable pathway leverages our key novel finding: confidence prediction via gaze transitions, a signal uniquely available in simultaneous pairwise comparison and distinct from prior work on sequential text-based evaluation. Gaze-derived confidence could weight annotations during reward model training such that high-confidence judgments exert greater influence, filter uncertain trials from training data, or flag ambiguous cases for secondary review. Additionally, within-image attentional patterns could support post-hoc analysis to identify which visual features drive preferences and where participants attend when deciding. Scaling these approaches beyond the laboratory presents challenges, though advances in webcam-based eye-tracking and wearable devices continue to improve accessibility and accuracy. 

Together, these findings provide a proof-of-concept that implicit gaze signals contain information about both what people prefer and how confident they are in those preferences, establishing a foundation for gaze-informed preference data collection in AI alignment.

\section{Privacy and Ethics Statement}
This study was approved by the Columbia University IRB, and all participants provided informed consent. Eye-tracking data were anonymized and unlinkable to individuals. We acknowledge that gaze-based preference prediction could potentially be applied without users' knowledge or consent, enabling covert inference of preferences for manipulative purposes. We emphasize that such applications would require explicit consent and transparent disclosure.

\paragraph{Acknowledgments.}
This work was supported by funding from the Army Research Laboratory’s STRONG Program (W911NF-19-2-0139, W911NF-19-2-0135, W911NF-21-2-0125), the Air Force Office of Scientific Research (FA9550-22-1-0337), a Vannevar Bush Faculty Fellowship from the US Department of Defense (N00014-20-1-2027) , and a Google Research Gift.

\bibliographystyle{unsrt}  
\bibliography{references}  

\end{document}